\newbox\grsign \setbox\grsign=\hbox{$>$} \newdimen\grdimen \grdimen=\ht\grsign
\newbox\simlessbox \newbox\simgreatbox
\setbox\simgreatbox=\hbox{\raise.5ex\hbox{$>$}\llap
     {\lower.5ex\hbox{$\sim$}}}\ht1=\grdimen\dp1=0pt
\setbox\simlessbox=\hbox{\raise.5ex\hbox{$<$}\llap
     {\lower.5ex\hbox{$\sim$}}}\ht2=\grdimen\dp2=0pt

\documentclass[10pt,preprint2]{aastex}

\slugcomment{}

\shorttitle{New Delta Scuti Stars}
\shortauthors{Sokoloski et al.}

\newcommand{\msun}{{M}_{\odot}}
\newcommand{\rsun}{{R}_{\odot}}

\begin{document}


\title{Four New $\delta$ Scuti Pulsators from a Variability Survey
of 131 Stars}


\author{J. L. Sokoloski}
\affil{Harvard-Smithsonian Center for Astrophysics,
    Cambridge, MA 02138}
\email{jsokoloski@cfa.harvard.edu}

\author{Lars Bildsten}
\affil{Institute for Theoretical Physics, Kohn Hall, University of California,
Santa Barbara, CA 93106}

\and

\author{R. Chornock and Alexei V. Filippenko}
\affil{Department of Astronomy, University of California,
    Berkeley, CA 94720-3411}



\begin{abstract}
In a photometric variability survey of 131 stars with $B \lesssim 14$
mag, we have found four new $\delta$ Scuti stars.  We were sensitive
to oscillation amplitudes as low as a few mmag.  The detection rate of
short-period ($P < 0.1$ d) pulsating variable stars, which may be
relevant for planned large variability surveys such as GAIA, POI, and
even the LSST, was therefore $3\%$.  All four new variable stars have
low pulsation amplitude (tens of mmag), and one has a very short
period (0.0198 d).  This star is one of the fastest known $\delta$
Scuti pulsators.  The short period of this variable star makes it
observationally tractable, and it may therefore be a particularly good
candidate for asteroseismological studies.  All four new variable
stars will add to the cadre of low-amplitude and relatively
short-period $\delta$ Scuti stars that are potentially useful for
learning about the structure of stars on or near the main sequence,
slightly more massive than the Sun.
\end{abstract}




\section{Introduction}

$\delta$ Scuti stars are pulsating A-type or F-type stars with mass $M 
\approx 1.5 - 2.5~\msun$ that lie in the classical Cepheid instability
strip.  They are usually on or near the main sequence 
(luminosity class V - III; Guzik 1997),
and they pulsate radially, as well as non-radially in pressure (p) and
sometimes gravity (g) modes.  The pulsation instability is driven by
opacity changes related primarily to the second ionization of He
($\kappa$-mechanism; \nocite{gs96} Gautschy \& Saio 1996).  The
photometric oscillations of the high-amplitude $\delta$ Scuti stars
(HADS), which are defined as those with $\Delta V > 0.3$ mag
\citep{breginbook00}, are generally dominated by the radial
fundamental mode.  In contrast, the variations seen in the
lower-amplitude $\delta$ Scuti stars consist of multiple non-radial
and radial p- (and g-) mode oscillations.  Most $\delta$ Scuti stars
are Population I objects, some of which show chemical peculiarities
(Rodr\'{\i}guez \& Breger 2001; Morgan \& Abt 1972).

In the course of a search for rapid optical variations in symbiotic
binary stars, Sokoloski, Bildsten, \& Ho (2001; hereafter SBH)
observed 131 stars in the fields of 35 symbiotic stars, primarily in
the northern hemisphere.  We describe here the serendipitous discovery
of variability in 4 of these stars, referred to as SBCF 1-4.  One new
variable object, SBCF 1, is quite clearly a $\delta$ Scuti star, and
given their oscillation periods and colors, the three others are
probably $\delta$ Scuti stars as well.  A fifth star from the sample,
SAO 31628, was found to be an eclipsing binary, and the observations
and ephemeris for this system are presented in
\cite{ss00}.
SBCF 1 has a 0.0198-d pulsation period, which is shorter than all but
3 of the 636 $\delta$ Scuti stars (with periods ranging from 0.0156 to
0.2878 d) listed in the catalogue of \cite{rod00}.  Very rapidly
pulsating main-sequence $\delta$ Scuti stars, such as this one, can
potentially be useful targets for study and mode identification, since
good frequency coverage can be obtained with a smaller temporal baseline
than for the more slowly pulsating systems.  Analysis of the pulsation
frequencies of $\delta$ Scuti stars provides an opportunity to test
models of stellar structure and evolution, estimate the ages of stars,
and determine the distances to stars through the use of a
period-luminosity(-color-metallicity) relationship.  To perform this
asteroseismology, however, it is necessary to identify the oscillation
modes (i.e., the spherical harmonic degree $l$, the azimuthal order
$m$, the radial degree (overtone) $n$ or $k$, and whether the mode is
a p mode, g mode, or mixed).

A new generation of survey experiments are being planned or are under
way that are geared specifically to the broad study of variable stars
or asteroseismology, such as the GAIA space mission (Perryman et
al. 2001), STARE (Stellar Astrophysics \& Research on Exoplanets;
Brown \& Kolinski 1999), ROTSE (the Robotic Optical Transient Search
Experiment; see Akerlof et al. 2000), ASAS (the All Sky Automated
Survey; Pojma\'{n}ski 2000), the Hungarian Automated Telescope (Bakos
2001), the Faint Sky Variability Survey (Groot et al. 1999), the
Panoramic Optical Imager Project (POI; Kaiser et al. 2002), and
eventually the Large-aperture Synoptic Survey Telescope (LSST; Tyson
\& Angel 2001).  Large
numbers of new $\delta$ Scuti variables are expected to be found as a
result of these projects
\citep{ec00}.  Estimates, however, are often based upon previous large
surveys, such as Hipparcos, that are not sensitive to short-period
pulsations, so there is much uncertainty in these predictions for new
$\delta$ Scuti stars.  Although our survey was very small compared to
Hipparcos, we were sensitive to shorter-period pulsations, and
so this work provides a glimpse of the variability fraction of
short-period variable stars outside of open clusters (where most
previous small surveys with similar sensitivity have been done).

We briefly describe our survey in
\S\ref{sec:observations}. The observations and results for the
individual new variable stars are presented in \S\ref{sec:sbcf1}
through \S\ref{sec:sbcf4}, and we discuss our conclusions in
\S\ref{sec:discussion}.

\section{Observations}\label{sec:observations}

Between 1997 and 1999, we performed a variability survey at UCO/Lick
Observatory, with the 1-m Nickel telescope, in order to search for
rapid variations in symbiotic binary stars.  Details of these
observations and the aperture photometry can be found in SBH.  
We were sensitive to oscillations with periods in the range of a few
minutes to several hours, and amplitudes down to a few mmag.  Only stars with $B
\lesssim 14$ mag were considered.
The periods for the pulsating variable stars discovered in this way
range from approximately 0.02 to 0.1 d, and the full $B$-band
oscillation amplitudes ranged from 10 to 75 mmag.

On 2001 November 18 (UT is used throughout this paper), to follow up on one
of the variable star discoveries, we obtained a moderate-resolution
spectrum of SBCF 1 using the Echellette Spectrograph and Imager (ESI;
\nocite{shei00}Sheinis et al. 2000) on the 10-m Keck II telescope.  A single 300-s
exposure in good conditions with a 1$\arcsec$ slit gave a spectral
resolution of $\sim 75$ km s$^{-1}$ over the range 3950-10400~\AA.
In order to get better spectral coverage in the near-ultraviolet, we
re-observed the object on 2002 January 18 for 300 s with the Low
Resolution Imaging Spectrometer (LRIS; \nocite{oke95}Oke et al. 1995)
mounted on the 10-m Keck I telescope.  The $0.7''$ slit provided a
resolution of 3.3~\AA\, over a wavelength range of 3350-5380~\AA\, on
the blue arm of the spectrograph and 2.0~\AA\, over 5700-7000~\AA\, on
the red arm.  The slit was aligned with the parallactic angle to
minimize the effects of atmospheric dispersion \citep{fil82} for both
observations.  Each two-dimensional exposure was processed using the
standard CCD reduction packages in IRAF\footnote{IRAF is distributed
by the National Optical Astronomy Observatories, which are operated by
the Association of Universities for Research in Astronomy, Inc., under
cooperative agreement with the National Science Foundation.}. Flux
calibration and removal of telluric absorption features \citep{waho88}
were accomplished in IDL using our own tasks (\nocite{math00}Matheson
et al. 2000).  The individual ESI orders were reduced separately and
then rebinned to a common wavelength scale before being added
together.

\begin{deluxetable}{lcrrccccc}
\tabletypesize{\footnotesize}
\tablecaption{New Variable Stars -- Basic Data and Observation Log\label{tab:tbl-1}}
\tablewidth{0pt}
\tablehead{
\colhead{Star} &\colhead{Catalog \&} & \colhead{RA\tablenotemark{a}} &
\colhead{Dec\tablenotemark{a}} & \colhead{$l$} & \colhead{$b$} & \colhead{$m_B$} &  
\colhead{Obs. Date(s)} & \colhead{Obs. Length} \\
\colhead{} & \colhead{Number} &\colhead{(h m s)} &\colhead{($^\circ$ $\arcmin$ $\arcsec$)
} & \colhead{$^\circ$} & \colhead{$^\circ$} & \colhead{}
&\colhead{(mm/dd/yy)} & \colhead{(hours)} } 
\startdata
SBCF 1 & USNO-A2.0  & 01 36 15.8 & +54 14 07 & 129.52 & -8.06 &
13.3$^b$,$13.8^d$ & 01/23/98 & 3.9 \\
& 1425-02166463 & & & & & & 09/15/98 & 4.5 \\
SBCF 2 & GSC 02137-00847 & 19 24 03.7& +29 40 32 & 62.95 & +6.66 &
11.4$^c$,11.4$^d$& 04/06/97 & 2.5  \\ 
       & & & & & & & 07/11/97 & 4.1 \\
       & & & & & & & 07/01/98 & 5.6 \\
SBCF 3 & GSC 01619-02513 & 19 45 34.5 & +18 40 30 & 55.67 & -2.94 &
12.8$^b$,$12.5^d$ & 06/29/98 & 2.3 \\ 
SBCF 4 & GSC 03645-01592 & 23 33 24.1 & +48 45 38 & 109.92 & -12.13 & 12.2$^b$,11.6$^c$, 
$12.1^d$& 08/30/97 & 6.5 \\
\enddata

\tablenotetext{a}{Coordinates are J2000.  Coordinates for SBCF 1 
are from the USNO-A2.0 catalogue, for SBCF 2 and SBCF 4 are
from the Tycho-2 main catalogue \citep{hog00}, for SBCF 3 are from the
Guide Star Catalogue.}
\tablecomments{b. USNO-A2.0 catalogue; c. Tycho
catalogue; d. From comparison with nearby symbiotic star.}

\end{deluxetable}

\section{SBCF 1: A Very Short Period Pulsator}\label{sec:sbcf1}

\begin{figure}[t]
\plotone{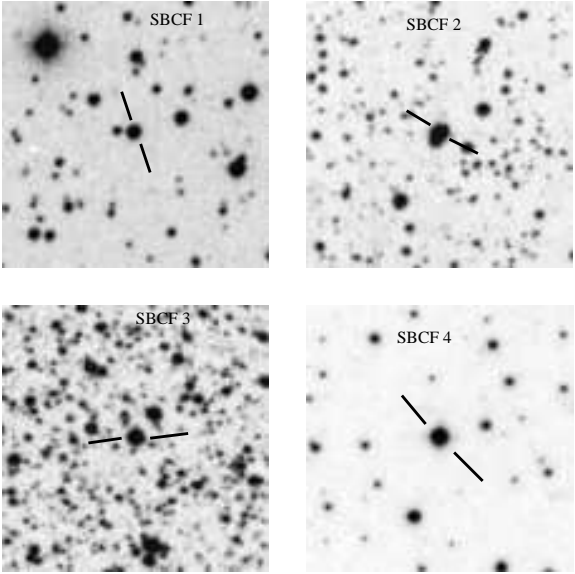}
\caption[]{\footnotesize Finder charts for SBCF 1-4.
North is up, and east is to the left. The fields are $3\arcmin
\times 3\arcmin$, and the coordinates for each object are listed in Table
\ref{tab:tbl-1}.  SBCF 2 is the more northern and brighter of the two
stars whose images are blended on the upper right
chart. (These charts are from the STScI Digitized Sky Surveys, which were
produced at the Space Telescope Science Institute under U.S. Government
grant NAG W-2166.)\label{fig:charts}}
\end{figure}

In the field of the symbiotic system AX Persei, we found an
oscillating star with a dominant period of 0.0198 d, and a total
peak-to-peak amplitude of 35 mmag.  This star does not have an SAO or
HD number (in the USNO-A2.0 catalogue, it is designated
1425-02166463), so we shall follow a recommendation from the IAU
(L. Dickel, private communication) and refer to it as SBCF 1.  It is
roughly 0.8 mag fainter than AX Per (in quiescence) in both of our
observations.  From \cite{mun92}, the $B$ magnitude of AX Per in
quiescence is 13.0, so the $B$ magnitude of SBCF 1 is approximately
13.8 (see Table \ref{tab:tbl-1} for the average magnitudes of the new
variable stars).  There is no object in the SIMBAD database at this
position.  The finder charts for this and the other new $\delta$
Scuti stars are shown in Figure
\ref{fig:charts},
and the coordinates of all objects are listed in Table
\ref{tab:tbl-1}.

\begin{figure}[t]
\plotone{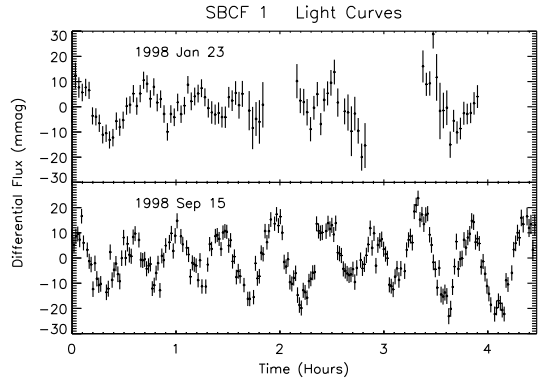}
\caption[Light Curves for SBCF 1, in the Field of AX
Per]{\footnotesize Differential $B$-band light curves from our two
observations of SBCF 1.  The first observation was severely affected
by clouds, but the 0.0198-d oscillation is evident in both
observations.  In the 1998 September 15 light curve, the oscillation
amplitude appears variable, indicating the likely presence of more
than one oscillation mode.  In fact, a second oscillation with a
period of 0.0265 d is detected. The time between data points is 118 s
and 66 s for the two observations, respectively.
\label{fig:sbcf1lcs}}
\end{figure}

We show the light curves from our two observations of SBCF 1 in Figure
\ref{fig:sbcf1lcs}.  The first observation, on 1998 January 23, was done
under cloudy conditions.  Consequently, the light curve from this
night has data gaps and is of generally poor quality.  During the
second observation, however, on 1998 Septempber 15, the conditions were good,
and the signal is clear.  Furthermore, the 1998 September 15 observation was
performed with a more sensitive CCD, and so we were able to obtain
higher temporal resolution.  The light curve from this date reveals that
the oscillation amplitude is variable, and that the object is
therefore multi-periodic.  In fact, at least two oscillation periods
can be identified using the phase dispersion minimization method
(Stellingwerf 1978), one at $0.0198 \pm 0.0001$ d, and another at
$0.0265 \pm 0.0006$ d.  The 0.0198-d oscillation has an amplitude of
24 mmag, as determined by folding the light curve with that period,
and the 0.0265-d oscillation has a smaller amplitude of roughly 10
mmag peak-to-peak, using the same folding method.  Both oscillations
have near-sinusoidal profiles.  Folding the light curve from January,
and again using the phase dispersion minimization method, we recover
an oscillation with period $P = 0.0197 \pm 0.0002$ d, consistent with
the stronger of the two oscillations detected in 1998 September.  The
parameters for SBCF 1 and the other new variable stars are listed in
Table \ref{tab:tbl-2}.

\begin{figure}[t]
\plotone{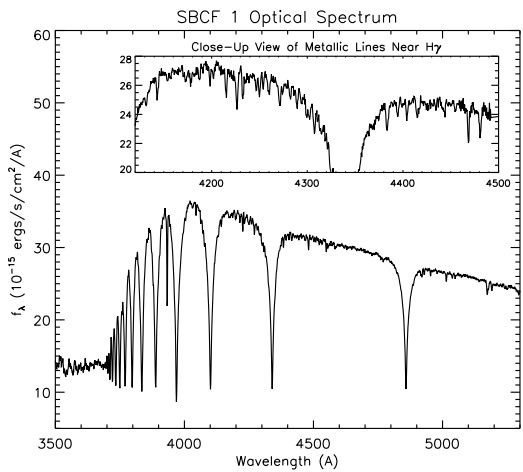}  
\caption[Spectrum of SBCF 1, in the field of AX Per]{\footnotesize Optical spectra of
SBCF 1.  The resolution of the full spectrum, taken with the Keck I
telescope on 2002 January 18, is 3.3 \AA.  The resolution of the
spectrum in the inset, taken with the Keck II telescope 2001 November
18, is 75 km s$^{-1}$ (or 1~\AA\, near $H
\gamma$). \label{fig:sbcf1spec}}
\end{figure}

The spectrum of SBCF 1 is shown in Figure \ref{fig:sbcf1spec}.  The
strong Balmer series and lack of He lines indicate that this is an
A-type star.  The strength of the Ca\,II K $\lambda$3934 feature, the
Balmer line equivalent widths, and visual comparison of the metallic
lines with spectral standards in the online digital spectral atlas of
R. O. Gray (http://nedwww.ipac.caltech.edu /level5/Gray/frames.html)
all indicate spectral type A2.  SBCF 1 is therefore a very short
period $\delta$ Scuti star.
In fact, the dominant period of 0.0198 d is shorter than that of any
of the 192 $\delta$ Scuti stars listed in the catalog of
\citet{lrrg90}, and all but 3 of 636 the stars listed in the catalog of
\cite{rod00}.

\begin{deluxetable}{lcccc}
\tablecaption{New Variable Star Parameters\label{tab:tbl-2}}
\tablewidth{0pt}
\tablehead{
\colhead{Star} &   
\colhead{Period}   & \colhead{$B$ Amplitude} &
\colhead{Spec. Type}  &
\colhead{Comments} \\
\colhead{} &\colhead{(d)} &\colhead{(full, mmag)} &\colhead{or color} &\colhead{}
}
\startdata
SBCF 1 & $0.0198 \pm 0.0001$ & 24 & A2V & Multiple
modes\\
 & $0.0265 \pm 0.0006$ & 10 & & \\
SBCF 2 & $0.06\pm 0.005$ & 36,55,74 & $B-V=0.4^a$ & Multiple modes\\
 & $0.094\pm 0.001$ & 35 &  & \\
SBCF 3 & $0.088 \pm 0.001$ & 30 & $B-R=0.4^b$ & \\
SBCF 4 & $0.080 \pm 0.005$ & 40 & $B-V=0.3^a$ & Poor sine fit\\
\enddata
\tablecomments{a. Tycho-2 main catalogue, b. USNO-A2.0 catalogue.}
\end{deluxetable}

To investigate the luminosity class, we examined the Balmer-line
profiles and the equivalent width of the O\,I $\lambda$7774 line.  We
estimated the rotational line broadening by looking at the
half-intensity width of the Mg\,II $\lambda$4481 line in our first,
higher resolution Keck spectrum.
Comparing the $\lambda$4481 half-intensity value of 2.8 \AA\, to the
half-intensity widths of the standard stars with similar spectral
types listed in Slettebak et al. (1975), we find $v \sin i = 115 \pm
5$ km s$^{-1}$.  
The Balmer-line profiles indicate that the star is not significantly
evolved.  The small O\,I $\lambda$7774 equivalent width of 0.7--0.8
\AA\, confirms that SBCF 1 is a main-sequence star (Jaschek \& Jaschek
1987), and the spectral type is thus A2V.  Luminosity class V is in
fact expected, since $\delta$ Scuti stars with low amplitudes and
short periods are generally not evolved (Rodr\'{\i}guez \& Breger
2001).  Furthermore, the measured value of $v \sin i$ for SBCF 1 is
very close to the average value for early A main-sequence stars
(Jaschek \& Jaschek 1987).  From the Schlegel, Finkbeiner, \& Davis
(1998) dust map, the reddening toward SBCF 1 is $E(B-V) = 0.23$ mag.
Taking the absolute magnitude $M_V = +1.3$ mag for an A2 main-sequence
star (Schmidt-Kaler 1982), $A_V = 0.7$ mag, and the measured apparent
magnitude, we find that SBCF 1 is roughly 2 kpc away.  This distance
estimate justifies the assumption that SBCF 1 is behind the full
column of extinction, given its Galactic latitude of $-8^\circ$.

For a spectral type A2V,
we can estimate the pulsation constant 
\begin{eqnarray*}
Q = \Pi (M/\msun)^{1/2} (R/\rsun)^{-3/2}
\end{eqnarray*}
 (where $\Pi$ is the pulsation period, $R$ is the
stellar radius, and $M$ is the stellar mass; Christensen-Dalsgaard
1993).  Using the expression $\log Q = -6.454 -\log f + 0.5 \log g +
0.1 M_{bol} + \log T_{eff}$ (where $f$ is the pulsation frequency in
cycles/day, $M_{bol}$ is the bolometric magnitude, and $T_{eff}$ is
the effective temperature; Breger \& Bregman 1975), and taking $\log
[g/({\rm cm\,s}^{-2})] = 4.2$,
$M_{bol} = +1.1$ mag, and $\log (T_{eff}/{\rm K}) = 3.953$
(Schmidt-Kaler 1982), we find $Q
\approx 0.010$ for the 0.0198-d oscillation, and $Q \approx 0.014$ for 
the 0.0265-d oscillation.  These values are obviously only
approximate, since $g$, $M_{bol}$, and $T_{eff}$ are all very
uncertain, but they are nonetheless interesting.  $Q$ is low and
therefore suggests high-degree non-radial or high-order radial
oscillations.  Non-radial oscillations with high $l$ value are not
expected to have the relatively large amplitude that we observe
(Balona \& Dziembowski 1999), so the oscillations are probably
high-overtone radial pulsators.  High-overtone, low spherical harmonic
degree pulsations (e.g., $l=1$ or $l=2$) are also a possibility
(J. Guzik, private communication).  Koen et al. (1999) presented a
similar argument for high-order radial oscillations in another very
short period $\delta$ Scuti star, HD 23194 ($P$ = 0.0204 d).

\section{SBCF 2 (GSC 02137-00847)}\label{sec:sbcf2} 

\begin{figure}[t]
\plotone{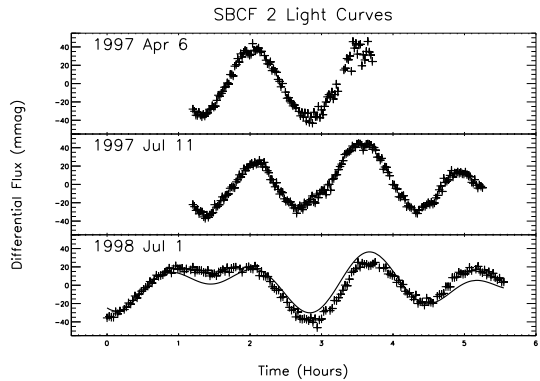}
\caption[]{\footnotesize Differential $B$-band light curves for SBCF
2, in the field of BF Cygni. Superposed on the third night of data is
the fit with 2 sine components ($P_1=0.094$ d, $P_2 = 0.06$
d). \label{fig:sbcf2lcs}}
\end{figure}

Contained in the field of the symbiotic star BF Cygni is the
star SBCF 2
(= GSC 02137-00847 = TYC 2137 847 1).  There is no indication from the
SIMBAD database that this star is variable, but a quick look at the
three light curves in Figure
\ref{fig:sbcf2lcs} reveals very significant
oscillations.  The variations are not simply repeatable, as we would
expect for a binary, so they are likely due to stellar pulsation.
>From the light curves, it is also apparent that more than one
pulsation mode is present, and that the peak-to-peak amplitude in $B$
can reach more than 70 mmag.  The first light curve, from 1997 Apr 6,
can be fit with a sine curve with period 0.064 d and peak-to-peak
amplitude of 74 mmag.  The second and third light curves, from 1997
Jul 11 and 1998 Jul 1, each require two sine components.  From the
1997 Jul 11 observation, we find sinusoidal oscillations with $P =
0.06$ d (55 mmag peak-to-peak) and $P = 0.12$ d (32 mmag
peak-to-peak), but since the 0.012-d period is close to the length of
the latter observation, we consider the detection marginal.  We fit
the longest light curve, from 1998 July 1, with a double sine wave,
and find periods of 0.094 d (35 mmag peak-to-peak) and 0.060 d (36
mmag peak-to-peak).  Given the uncertainty in the measured periods, it
is possible that the 0.064-d oscillation on that day is consistent
with the 0.06-d oscillations detected on the other nights.  Longer
observations are clearly needed to determine the number of modes
present in this star, although it is clear that there are at least
two.  According to the General Catalogue of Variable Stars (Kholopov
et al.  1992), pulsating variable stars with periods in this vicinity
include $\delta$ Scuti stars, with periods in the range 0.01 to 0.2 d
(Rodr\'{\i}guez et al. 2000 put the upper limit at $0.3$ d, and list
the shortest period as 0.0156 d), and the SX Phe stars.  The SX Phe
stars, however, generally have amplitudes greater than 0.25 mag
(McNamara 1997), so SBCF 2 is unlikely to fall in this category.

From the Tycho-2 main catalogue \citep{hog00}, the magnitudes for this
object are $B_T = 11.537\pm 0.069$ mag and $V_T=11.092\pm0.076$
mag. Converting these values to the Johnson system (using the
expressions from the Tycho catalogue description), we find  $V=11.05
\pm 0.083$ mag, $B-V=0.38 \pm 0.09$ mag, and thus $B=11.43 \pm 0.12$ mag.
The reddening from \cite{sfd98} at the position of SBCF 2 is
$E(B-V)=0.287$ mag, so for this star, $-0.06 \leq (B-V)_0 \leq 0.47$
mag. This range of $(B-V)_0$ corresponds to an A- or F-type star,
which is exactly what we expect for a $\delta$ Scuti pulsator.
Therefore, given the oscillation period, light-curve shape, intrinsic
color $(B-V)_0$, and pulsation amplitude (which ranges from 3 mmag to
0.9 mag in $V$ for $\delta$ Scuti stars), we conclude that SBCF 2 is a
$\delta$ Scuti star.  Using absolute magnitudes and extinctions
corresponding to spectral types within the allowed range, the
calculated distance to SBCF 2 is at least 250 pc.

\section{SBCF 3 (GSC 01619-02513)}\label{sec:sbcf3} 

\begin{figure}[t]
\plotone{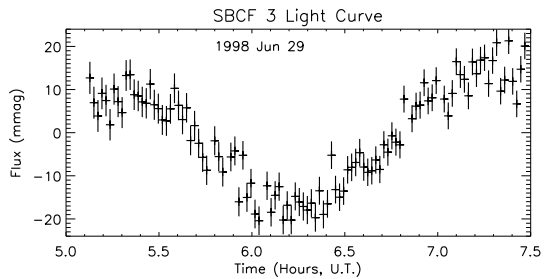}
\caption[]{\footnotesize Differential $B$-band light
curve for SBCF 3, in the field of symbiotic star AS 360.  The data
points are separated by 78 s. \label{fig:sbcf3lc}}
\end{figure}

In the field of the symbiotic star AS 360 (= QW Sge), 
the star SBCF 3 (= GSC 01619-02513) oscillated with a peak-to-peak
$B$-band amplitude of 30 mmag, and a period of 2.1 hr. 
We show the light curve for this star in Figure
\ref{fig:sbcf3lc}. There is no object in the SIMBAD database at this position.
In the USNO-A2.0 catalog, the $B$ magnitude of this star (USNO-A2.0
1050-14753932) is listed as 12.8, and the $R$ magnitude as 12.4,
giving a color for SBCF 3 of $B-R=0.4$ mag.
 The intrinsic color for SBCF 3 is therefore $(B-R)_0
\le 0.4$ mag;  
if it is a main-sequence star, SBCF 3 must therefore be spectral type
A or earlier.  Since an A-type star has a radius roughly two or more times
that of the Sun, such a star cannot fit within the Roche lobe of a
binary with either a 2-hr or 4-hr orbital period (the orbital period
of an ellipsoidal variable would be twice the photometric period).  To
fit within the Roche lobe of a 2-hr binary, a star would need $M < 0.3
\msun$.  The oscillation period of 2.1 hr and the restriction on
the spectral type indicate that SBCF 3 is likely to be a $\delta$
Scuti star.

\section{SBCF 4 (GSC 03645-01592)}\label{sec:sbcf4} 

\begin{figure}[t]
\plotone{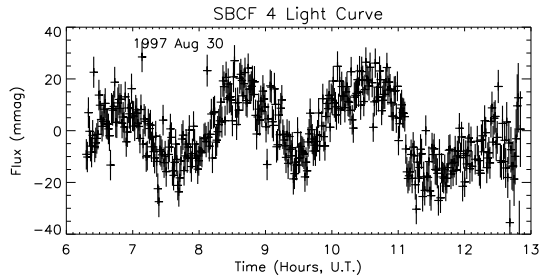}
\caption[]{\footnotesize Differential $B$-band light
curve for SBCF 4, in the field of Z And.  The data points are
separated by 58 s. \label{fig:sbcf4lc}}
\end{figure}

SBCF 4 (= GSC 03645-01592 = TYC 3645 1592 1) is the only star within
$7\arcmin$ of Z And whose brightness is within 2 or even 3 mag
of Z And.  
In the Tycho catalogue, photometry for GSC 03645-01592 is given as
$B_T = 11.64\pm 0.06$ mag and $V_T = 11.38\pm 0.076$ mag.  Converting
to Johnson magnitudes,
we find $B=11.6\pm 0.1$ and $V=11.4 \pm 0.1$ mag.  
Because of the interesting 28-minute oscillation in Z And (Sokoloski
\& Bildsten 1999), we observed this field many times.  The paucity of
suitable comparison stars, however, made high-quality light curves
difficult to produce.  An example, which was constructed using an
ensemble average of 2 very faint comparison stars, is shown in Figure
\ref{fig:sbcf4lc}.  The signal-to-noise ratio is lower than for the
other light curves, but variability at the level of about 40 mmag
peak-to-peak is evident.  The variability amplitude of this star was
not constant from observation to observation (i.e., month to month),
and on at least one occasion, the variations were undetectable.  When
variations were present, the oscillation period for SBCF 4 was roughly
2 hr.  With an intrinsic color $(B-V)_0 \le 0.3$ mag, the spectral
type of this star must be F0 or earlier.
As with SBCF 3, a star with such an early spectral type could not fit
within the Roche lobe of a binary with the observed (or twice the
observed) period.  Therefore, with an A- or early-F-type spectrum, and
an oscillation period of 2 hr, this star is an excellent candidate
$\delta$ Scuti star.

\section{Discussion}\label{sec:discussion}

We analyzed light curves for 131 stars with $B$ magnitude less than
14, that had a typical temporal resolution of 1 minute or shorter.
Out of these objects, we found four new pulsating variable stars.  The
short-period $\delta$ Scuti star SBCF 1 is potentially a good
candidate for further study, since mode identification might be
accomplished with less observing time than for the longer period
$\delta$ Scuti stars. SBCF 1 is not as bright as some of the other
$\delta$ Scutis that have been targets for serious asteroseismological
efforts, but with the availability of larger telescopes and more
sensitive CCDs, the opportunity to use a shorter temporal baseline,
and therefore less complex window function, becomes increasingly
important. For example, using differential CCD photometry and a 1-m
telescope, we were able to obtain the same level of photometric
precision (several mmag per point) for SBCF 1 as Viskum et al. (1998)
had in their study of FG Vir that resulted in the identification of
eight oscillation modes of that star.  In addition, unevolved, core
H-burning $\delta$ Scuti stars have pulsation spectra that are
theoretically predicted to be less dense than shell H-burning, evolved
$\delta$ Scuti stars (Michel et al. 2000).  Asteroseismological
studies of this object could thus help achieve broad goals such as
testing theories of stellar structure and evolution, and extending the
period-luminosity relation to low-amplitude pulsating stars (Petersen
\& Christensen-Dalsgaard 1999).

Nearly all currently known $\delta$ Scuti stars with oscillation
amplitudes below 20 mmag or periods below 0.05 d have been found by
individual observers or small collaborations
(\nocite{rod01}Rodr\'{\i}guez \& Breger 2001).  Therefore, in contrast
to the longer period variables that are easily detected by, for example,
Hipparcos, MACHO, and OGLE, very little statistical information is
currently available about the prevalence of stellar variability in
this parameter range.  In fact, the studies which heretofore have been
most sensitive to rapid, low-amplitude oscillations have primarily
examined stars in open clusters, where the detection rate is a
function of cluster age\footnote{Some other unknown factor also
affects the detection rate in clusters.  No $\delta$ Scutis at all
were found out of 68 stars in the instability strip in NGC 7226
(Viskum et al. 1997).} (since a different amount of the isochrone will
pass through the instability strip for different ages).  Therefore,
small surveys such as this one, with the ability to detect variations
on time scales as short as a few minutes, can provide information
that, when considered with large survey results, should improve the
predictions of the number of variable stars that will be found with
the new generation of searches.

Our discovery rate for low-amplitude pulsators of 3\% is higher than
the rates that were found with Hipparcos (1\%; Eyer 1999), OGLE (0.04
- 0.08\%; Udalski et al. 1994, 1995a,b; Eyer 1999), or MACHO ($<$
0.032\%; Cook 1997).  The difference between these results could be
due to the small-number statistics for our survey, or to our better
sensitivity in the more highly populated $\delta$ Scuti period and
amplitude parameter ranges.  The number of $\delta$ Scuti stars found
with Hipparcos dropped off below $\Delta V \approx 40$ mmag
\citep{rod01}, and the OGLE and MACHO surveys are not sensitive to
pulsations with amplitude less than 0.1 mag (Rodr{\'i}guez et
al. 2001).

The typical periods of $\delta$ Scuti stars
may be even more relevant for explaining our higher detection rate.
The peak of the distribution of all $\delta$ Scuti stars with period
lies between 0.05 and 0.1 d, and more than 10\% of known
$\delta$ Scuti stars have pulsation periods less than 0.05 d
\citep{rod01}.  We had excellent sensitivity to variables with periods
in this range.  Hipparcos, on the other hand, because of its sampling,
discovered very few $\delta$ Scuti stars with pulsation period less
than 0.1 day, and practically none with pulsation period less than
0.05 day.
To put our results in the context of other surveys, ASAS looks for
variables at a cadence of one day, and they have detected pulsating
stars with periods only down to 0.13 d.
The ROTSE all-sky survey found a firm lower limit of 0.2\% for the
variability fraction in the first analysis of part of their survey,
but they are insensitive to oscillations with amplitude less than 0.1
mag (in addition to some period limitations from obtaining data points
only twice per day; Akerlof et al. 2000).

\acknowledgments

\begin{footnotesize}
We would like to thank to Dimitar Sasselov, Scott Kenyon, and Yanqin
Wu for useful discussions, as well as Joyce Guzik, Paul Bradley, and a
very thourough anonymous referee for their critical reading of the
manuscript.  We are also grateful to Jason Aufdenberg, Bob Kurucz, and
Anne Bragg for assistance with the initial spectral analysis of SBCF
1. We acknowledge the assistance of the staffs of the Lick and Keck
Observatories.  The W. M. Keck Observatory is operated as a scientific
partnership among the California Institute of Technology, the
University of California, and NASA; it was made possible by the
generous financial support of the W. M. Keck Foundation.  This
research was supported by NASA via grant NAG 5-8658 and by the NSF
under grants PHY-9907949, AST-0196422, AST-9987938, and INT-9902665.
L.B. is a Cottrell Scholar of the Research Corporation.  and A.V.F. is
a Guggenheim Foundation Fellow.
\end{footnotesize}

\begin{footnotesize}

\end{footnotesize}


\begin{thebibliography}{}
\bibitem[Akerlof et al.(2000)]{ake00}Akerlof, C., et al. 2000, AJ, 119, 
1901
\bibitem[Bakos(2001)]{bak01}Bakos, G. 2001, Pub. Ast. Dept.of the
E{\"o}tv{\"o}s Uni. No. 11, Proceedings of the National Postgraduate Reuenion
in Astronomy \& Astrophysics, 2000, 107
\bibitem[Balona \& Dziembowski(1999)]{bd99}Balona, L. A., \&
Dziembowski, W. A. 1999, MNRAS, 309, 221
\bibitem[Breger(2000)]{breginbook00}Breger, M. 2000, in Delta Scuti and
Related Stars, ed. M. Breger \& M. H. Montgomery (San Francisco:
ASP, Conf. Ser. 210), 3
\bibitem[Breger \& Bregman(1975)]{bb75}Breger, M., \& Bregman,
J. N. 1975, ApJ, 200, 343
\bibitem[Brown \& Kolinski(1999)]{bkol99}Brown, T., \& Kolinski,
D. 1999, http://www.hao.ucar.edu/ \\public/research/stare/stare.html
\bibitem[Christensen-Dalsgaard(1993)]{cd93}Christensen-Dalsgaard, J.
1993, in Inside the Stars, ed. W. W. Weiss \& A. Baglin (San
Francisco: ASP, Conf. Ser. 40), 483
\bibitem[Cook(1997)]{cook97}Cook, K. 1997, in Variable Stars and the
Astrophysical Returns of the Microlensing Surveys (12th IAP
colloquium), ed. R. Ferlet, P. Maillard, \& B. Raban (Paris: Editions
Fronti\`{e}res), 17
\bibitem[Eyer(1999)]{eyer99}Eyer, L. 1999, Baltic Astronomy, 8, 321
\bibitem[Eyer \& Cuypers(2000)]{ec00}Eyer, L., \& Cuypers, J. 2000, in 
The Impact of Large-Scale Surveys on Pulsating Star Research, ed. L. 
Szabados \& D. Kurtz (San Francisco: ASP, Conf. Ser. 203), 71
\bibitem[Filippenko(1982)]{fil82}Filippenko, A.~V. 1982, PASP, 94, 715
\bibitem[Gautschy \& Saio(1996)]{gs96}Gautschy, A., \& Saio, H. 1996, 
\araa, 34, 551
\bibitem[Groot et al.(1999)]{gro99}Groot, P., Everett, M., Howell, S., 
Vreeswijk, P., Huber, M., \& van Paradijs, J. 1999, AAS \#80.03
\bibitem[Guzik(1997)]{guzik97}Guzik, J. A. 1997, in New Eyes to See Inside
the Sun and Stars, ed.  F. Deubner, J. Christensen-Dalsgaard,
\& D. Kurtz (Dordrecht: Kluwer), 331
\bibitem[H{\o}g et al.(2000)]{hog00}H{\o}g, E., Fabricius, C., Kakarov,
V. V., Urban, S., Corbin, T., Wycoff, G., Bastian, U., Scwekendiek,
P., \& Wicenec, A. 2000, A\&A, 355, L27
\bibitem[Jascheck \& Jaschek(1987)]{jj87}Jaschek, C., \& Jaschek,
M. 1987, The Classification of Stars (Cambridge: Cambridge Univ. Press)
\bibitem[Kaiser et al.(2002)]{kai02}Kaiser, N., et al. 2002, see
http://poi.ifa.hawaii.edu/ 
\bibitem[Kholopov et al.(1992)]{kho92}Kholopov, P. N., Samus, N. N.,
Durlevich, O. V., Kazarovets, E. V., Kireeva, N. N., \& Tsvetkova,
T. M. 1992, General Catalogue of Variable Stars, 4th ed., vol IV,
Bull. Inf. CDS, 40, 15
\bibitem[Koen et al.(1999)]{koen99}Koen, C., Van Rooyen, R., Van Wyk,
F., \& Marang, F. 1999, MNRAS, 309, 1051 
\bibitem[L\'{o}pez de Coca et al.(1990)]{lrrg90}L\'{o}pez de Coca, P., 
Rolland, A., Rodr\'{\i}guez, E., \& Garrido, R. 1990, A\&ASS, 83, 51
\bibitem[Matheson et al.(2000)]{math00}Matheson, T., et al. 2000, AJ, 120, 1499 
\bibitem[Michel et al.(2000)]{mich00}Michel, E., et al. 2000, in The
Impact of Large-Scale Surveys on Pulsating Star Research,
ed. L. Szabados \& D. Kurtz (San Francisco: ASP, Conf. Ser. 203), 69
\bibitem[McNamara(1997)]{mcnam97}McNamara, D. 1997, PASP, 109, 1221
\bibitem[Morgan \& Abt(1972)]{ma72}Morgan, W. W., \& Abt, H. A. 1972,
AJ, 77, 35
\bibitem[Munari et al.(1992)]{mun92}Munari, U., et al. 1992, A\&ASS,
93, 383
\bibitem[Oke et al.(1995)]{oke95}Oke, J.~B., et al. 1995, PASP, 107, 375
\bibitem[Perryman et al.(2001)]{per01}Perryman, M. A. C., et al. 2001, 
A\&A, 369, 339
\bibitem[Petersen \& Christensen-Dalsgaard(1999)]{pcd99}Petersen,
J. O., \& Christensen-Dalsgaard, J. 1999, A\&A, 352, 547
\bibitem[Pojma\'{n}ski(2000)]{poj00}Pojma\'{n}ski, G. 2000,
AcA, 50, 177 
\bibitem[Rodr\'{\i}guez \& Breger(2001)]{rod01}Rodr\'{\i}guez, E., \&
Breger, M. 2001, A\&A, 366, 178 
\bibitem[Rodr\'{\i}guez et al.(2000)]{rod00}Rodr\'{\i}guez, E.,
L\'{o}pez-Gonz\'{a}lez, M. J., \& L\'{o}pez de Coca, P. 2000, A\&ASS,
144, 469
\bibitem[Schlegel et al.(1998)]{sfd98}Schlegel, D. J., Finkbeiner,
D. P., \& Davis, M. 1998, ApJ, 500, 525
\bibitem[Schmidt-Kaler(1982)]{s-k82}Schmidt-Kaler, T. 1982, in
Landolt-B{\"o}rnstein: Numerical Data and Functional Relationships in
Science and Technology, ed. K. Schaifers \& H. H. Voigt (Berlin:
Springer-Verlag)
\bibitem[Sheinis et al.(2000)]{shei00}Sheinis, A.~I., Miller, J.~S.,
Bolte, M., \& Sutin, B.~M. 2000, Proc. SPIE, 4008, 522 
\bibitem[Slettebak et al.(1975)]{slet75}Slettebak, A., Collins, G. W., 
Boyce, P. B., White, N. M., \& Parkinson, T. D. 1975 ApJS, 281, 137
\bibitem[Sokoloski \& Bildsten(1999)]{sb99}Sokoloski, J. L., \&
Bildsten, L. 1999, ApJ, 517, 919
\bibitem[Sokoloski et al.(2001)]{sbh01}Sokoloski, J. L., Bildsten,
L., \& Ho, W. C. G. 2001, MNRAS, 326, 553 (SBH)
\bibitem[Sokoloski \& Stone(2000)]{ss00}Sokoloski, J. L., \& Stone,
R. P. S. 2000, IBVS, 4983
\bibitem[Stellingwerf(1978)]{ste78}Stellingwerf, R. F. 1978, ApJ, 224,
953 
\bibitem[Tyson \& Angel(2001)]{tys01}Tyson, J. A., \& Angel, R. 2001, in
  The New Era of Wide Field Astronomy, ed. R. Clowes, A. Adamson, \& G.
  Bromage (San Francisco: ASP), 347
\bibitem[Udalski et al.(1994)]{udal94}Udalski, A., Kubiak, M.,
Szyma\'{n}ski, M., Kalu\.{z}ny, J., Mateo, M., \& Krzemi\'{n}ski,
W. 1994, Acta Astron. 44, 317
\bibitem[Viskum et al.(1997)]{visk97}Viskum, M., Hern\'{a}ndez, M. M.,
Belmonte, J. A., \& Frandsen, S. 1997, A\&A, 328, 158
\bibitem[Wade \& Horne(1988)]{waho88}Wade, R.~A., \& Horne,
K.~D. 1988, ApJ, 324, 411 
\end{thebibliography}
\end{document}